\documentclass[aps,prd,twocolumn,groupedaddress,showpacs]{revtex4}
\usepackage{graphicx}
\usepackage{dcolumn}
\usepackage{bm}
\usepackage{natbib}
\usepackage{multirow}
\usepackage{epsfig}
\usepackage{amsmath}
\usepackage{amssymb}

\newcommand{\beq}{\begin{equation}}
\newcommand{\eeq}{\end{equation}}
\newcommand{\beqa}{\begin{eqnarray}}
\newcommand{\eeqa}{\end{eqnarray}}

\newcommand{\be}{\begin{equation}}
\newcommand{\ee}{\end{equation}}

\renewcommand{\vec}[1]{{\bf #1}}

\begin{document}

\title{Relative velocity of dark matter and baryonic fluids and the formation of the first structures}

\author{Dmitriy Tseliakhovich}
\affiliation{California Institute of Technology, M/C 249-17, Pasadena, California 91125, USA}

\author{Christopher Hirata}
\affiliation{California Institute of Technology, M/C 350-17, Pasadena, California 91125, USA}

\date{August 25, 2010}

\begin{abstract}
At the time of recombination, baryons and photons decoupled and the sound speed in the baryonic fluid dropped from relativistic, $\sim c/\sqrt3$, to the thermal velocities of the hydrogen atoms, $\sim 2\times 10^{-5}c$.  This is less than the relative velocities of baryons and dark matter computed via linear perturbation theory, so we infer that there are supersonic coherent flows of the baryons relative to the underlying potential wells created by the dark matter.  As a result, the advection of small-scale perturbations (near the baryonic Jeans scale) by large-scale velocity flows is important for the formation of the first structures.  This effect involves a quadratic term in the cosmological perturbation theory equations and hence has not been included in studies based on linear perturbation theory.  We show that the relative motion suppresses the abundance of the first bound objects, even if one only investigates dark matter haloes, and leads to qualitative changes in their spatial distribution, such as introducing scale-dependent bias and stochasticity.  We further discuss the possible observable implications of this effect for high-redshift galaxy clustering and reionization.
\end{abstract}

\pacs{98.65.Dx, 98.80.Es}

\maketitle

\section{Introduction}

The early Universe was extremely homogeneous and isotropic, with small adiabatic density perturbations likely seeded during an epoch of 
inflation \cite{1981PhRvD..23..347G, 1982PhLB..108..389L, 1982PhRvL..48.1220A}.  The subsequent evolution of the Universe is well-described by a model containing 
baryons, cold dark matter (CDM), and a cosmological constant ($\Lambda$).  This inflationary $\Lambda$CDM paradigm, with only six parameters, is simultaneously 
consistent with a wide range of cosmological observables \cite{2007ApJS..170..377S, 2010arXiv1001.4538K}.

One of the key features of the $\Lambda$CDM scenario is the hierarchical formation of structure: since the matter power spectrum $\Delta^2(k)$ is an increasing 
function of wavenumber $k$, the smallest perturbations collapse first, followed by their assembly into larger and larger structures.  The formation of the first 
structures has recently become a major research area: aside from the intrinsic interest in understanding the first galaxies, these objects are believed to be 
responsible at least partially for the reionization of the intergalactic medium \cite{1994MNRAS.269..563F, 1996ApJ...472L..63O}, and they are sensitive to the small-scale power spectrum of the dark 
matter, which is a powerful probe of dark matter microphysics.

The evolution of density perturbations in the early Universe is generally described using linear perturbation theory, 
which treats overdensities and velocity fields as small
quantities and hence neglects second order terms.  Several previous works have extended this theory down to the post-recombination baryon Jeans scale \cite{1997PhRvD..56.7566Y, 1998ApJ...501..442Y}.
Interest in direct observations of the high-redshift Universe via absorption in the redshifted 21 cm line \cite{2004PhRvL..92u1301L}
has motivated more detailed investigation of the clustering of baryons during the epoch between recombination and reionization \cite{2005MNRAS.363L..36B, 2007MNRAS.377..667N},
including the entropy and ionization fluctuations in the baryons \cite{2005MNRAS.362.1047N, 2007PhRvD..76f3001L, 2007PhRvD..76h3005L}.  A deficiency of linear perturbation theory is that it does not 
describe the collapse of 
perturbations to 
form bound haloes, although analytical models such as the Press-Schechter formalism \cite{1974ApJ...187..425P, 1991ApJ...379..440B} are often used to estimate the halo mass function and clustering.
In order to go beyond linear perturbation theory, one may
use spherically symmetric Lagrangian hydrodynamic models \cite{1996ApJ...464..523H}
or
high-resolution 3D hydrodynamic simulations to follow the infall of baryons into the first haloes \cite{2002ApJ...564...23B, 2009MNRAS.399..369N}.  
Since it is inherently 
nonperturbative 
this approach can, with incorporation of appropriate chemistry and cooling processes, even be followed all the way to the formation of the first stars
\cite{2002Sci...295...93A, 2005ApJ...628L...5O, 2008Sci...321..669Y}. 

The principal purpose of this paper is to point out a new nonlinear effect in the growth of small-scale density perturbations that is active even at $z\sim 1000$.  The idea is that prior to 
recombination, the baryons are tightly coupled to the photons resulting in a standing acoustic wave pattern \cite{1970Ap&SS...7....3S}.  Modern linear perturbation theory treatments including the CDM 
\cite{1995ApJ...444..489H, 1995ApJ...455....7M} show a consequent relative velocity of the baryons and CDM since the latter does not suffer Thomson scattering and merely follows geodesics of the cosmic 
spacetime.  At the time of recombination, the root-mean-square (RMS) relative velocity is 30 km s$^{-1}$, and this is coherent over a scale of several Mpc comoving (the Silk damping scale 
\cite{1968ApJ...151..459S}).  When the baryons recombine and are no longer tied to the photons, their sound speed drops to 
$\approx6$ km s$^{-1}$, and hence there is a highly supersonic relative velocity between baryons and CDM.  This means that near the baryonic Jeans scale, perturbations in the baryons and CDM are advected 
relative to each other in less than a Hubble time (and hence less than their growth time).  This effect is investigated herein, and we find that it both suppresses the growth of small-scale structure, 
and leads to qualitatively new effects in the clustering of the first bound baryonic objects.

The suppression effect does not appear to have been present in previous analyses.  Since it results from the coupling 
of large-scale and small-scale modes, it is nonlinear and hence not present in 
linear perturbation theory.  Since the large-scale modes involved are associated with the acoustic oscillations of the photon-baryon fluid, they are not properly modeled by hydrodynamic simulations whose 
box size is smaller than the acoustic horizon ($\sim 140$ Mpc comoving).

Understanding of the physical processes that determine the collapse of the first dark matter halos and subsequent 
accumulation of baryonic matter in those halos is of paramount importance for interpretation of future data on 
reionization, high-redshift galaxies, and possibly dark matter substructure.
In the present paper we introduce the formalism and focus on the key features of the relative velocity 
effect, leaving a detailed study of various applications for future work.

This paper is organized as follows. In Section~\ref{sec:PT} we introduce the effect and calculate the Mach number of the relative motion of dark matter and baryonic 
fluids at the time of recombination. We show that accounting for the relative motion leads to a suppression of the matter power spectrum near the baryon Jeans scale.
In Section~\ref{sec:sim} we compute the 
abundance and clustering properties of the first haloes taking account of the relative motion.  The treatment is simple (it uses the linear Gaussian random field model for the large-scale density and 
velocity perturbations in cells of size a few Mpc, and then uses an analytic model to compute the density of small haloes in each cell), but we believe it should capture the qualitative result of the 
relative motion effect.  We briefly summarize our results and outline possible future work in Section~\ref{sec:conc}.

The numerical results and plots shown in this paper assume a cosmology with present-day baryon density $\Omega_{\rm b,0}=0.044$, CDM density $\Omega_{\rm c,0}=0.226$, dark energy density $\Omega_{\rm 
\Lambda,0}=0.73$, Hubble constant $H_0=71$ km$\,$s$\,$Mpc$^{-1}$, and adiabatic primordial perturbations of variance $\Delta^2_\zeta=2.42\times 10^{-9}$.


\section{Growth of small-scale structure including relative velocity of baryons and CDM}
\label{sec:PT}

Before recombination, baryons are tightly coupled to photons via Thomson scattering and the sound speed is that 
appropriate for a radiation-dominated plasma, $\sim c/\sqrt{3}$.  Perturbations in the CDM component can grow, however, 
because the CDM experiences no drag against the radiation. As the universe expands and cools electrons recombine with 
protons and the universe becomes transparent \cite{1968ApJ...153....1P, 1968ZhETF..55..278Z}. This also leads to a 
kinematic decoupling of the baryons from the radiation, so that the baryons can fall into the potential wells created 
by the CDM. The effective redshift of decoupling is $z_{\rm dec} \approx 1020$, which is slightly later than the 
surface of last scattering for microwave background photons because the baryons have lower inertia than the photons 
during this epoch \cite{1998ApJ...496..605E}.

\subsection{Basic setup}

In the post-recombination gas, the baryonic sound speed is
\beq
c_{\rm s} = \sqrt{\frac{\gamma k T_{\rm b}}{\mu m_{\rm H}}},
\eeq
where $\gamma = 5/3$ for an ideal monatomic gas, $\mu = 1.22$ is the mean molecular weight including a helium mass fraction of 0.24, $m_{\rm H}$ is the mass of the 
hydrogen atom, 
and $T_{\rm b}$ is the kinetic temperature of the baryons.  Here $T_{\rm b}$ is determined by a competition between adiabatic cooling and Compton heating from the 
CMB; we obtain it
using the {\sc Recfast} code \cite{1999ApJ...523L...1S, 2000ApJS..128..407S} and 
parametrize it as:
\beq
T_{\rm b}(a) = \frac{T_{\rm CMB,0}}{a}\left[ 1 + \frac{a/a_1}{1+(a_2/a)^{3/2}} \right]^{-1},
\eeq
with $a_1 = 1/119$, $a_2 = 1/115$, and $T_{\rm CMB,0} = 2.726 \ $K. 

While the baryonic velocity drops precipitously during recombination dark matter velocity remains unaffected and after recombination dark matter motion with respect 
to baryons become significant. The relative velocity can be written as:
\beq
{\bf v}_{\rm bc}({\bf k}) = \frac{\hat{\bf k}}{ik}[\theta_{\rm b}({\bf k}) - \theta_{\rm c}({\bf k})],
\eeq
where $\hat{\bf k}$ is a unit vector in the direction of ${\bf k}$, and $\theta \equiv a^{-1}\nabla\cdot{\bf v}$ is the velocity divergence.

\begin{figure}
\includegraphics[width=3.4in]{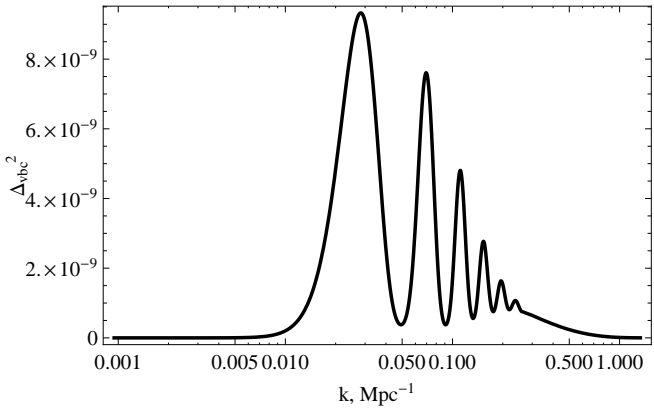}
\caption{\label{fig:cohFig}The coherence scale of $\vec{v}_{\rm bc}$ is determined by the range of scales over which $\Delta^2_{\rm vbc}(k)$ is non-zero. Here we plot $\Delta^2_{\rm vbc}(k)$, the 
variance
of the relative velocity perturbation per $\ln k$, as a function of wavenumber $k$. The power spectrum drops rapidly at $k>0.5\,$Mpc$^{-1}$, indicating that the relative velocity is coherent over scales 
smaller than a few Mpc comoving.}
\end{figure}

The variance of this relative velocity is
\beqa \label{eq:vbc}
\langle v_{\rm bc}^2(\vec{x}) \rangle &=& \int \frac{dk}{k} \Delta_\zeta^2(k)\left[ \frac{\theta_b (k) - \theta_c (k) }{k}\right]^2
\nonumber \\
&=& \int \frac{dk}{k} \Delta^2_{\rm vbc}(k),
\eeqa
where $\Delta_\zeta^2(k) = 2.42\times10^{-9}$ is the initial curvature perturbation variance per $\ln k$ 
\cite{2009ApJS..180..306D}. Integration of Eq.~(\ref{eq:vbc}) 
at the time of recombination ($z_{rec} = 1020$) shows 
that dark matter moves relative to the baryons with rms velocities $\sim30\,$km$\,$s$^{-1}$  corresponding to a Mach number of ${\cal M}\equiv v_{\rm bc}/c_{\rm 
s}\sim 5$. This supersonic relative motion allows baryons 
to advect out of the dark matter potential wells and significantly suppresses the growth of structure at wave numbers higher than
\beq
k_{\rm vbc} \equiv \left.\frac{aH}{\langle v_{\rm bc}^2\rangle^{1/2}}\right|_{\rm dec} 
= \frac{k_{\rm J}}{\cal M}
\sim 40\,{\rm Mpc}^{-1}.
\label{eq:kvbcdef}
\eeq

The relative contributions to $v_{\rm bc}$ from different scales are shown in Figure~\ref{fig:cohFig}.  One can clearly see that there is no contribution to the 
relative velocity from the 
largest scales, which were outside the sound horizon at the time of decoupling, and that the dominant contribution arises from the acoustic oscillation regime, 
which have typical velocities of a few times $c\Delta_\zeta$ and suffer no Hubble damping at early epochs when
$\rho_{\gamma}\gg\rho_{\rm b}$ \cite{1995ApJ...444..489H}.  At the smallest 
scales, the acoustic oscillations in the baryons are damped by photon diffusion, and the CDM velocities are suppressed by Hubble drag during the radiation era.  Thus 
we expect that ${\bf v}_{\rm bc}$ contains contributions from scales ranging from the Silk damping length up to the sound horizon.  This leads us to the conclusion 
that there is a {\em separation of scales}: the scales at which the first baryonic objects will form ($\sim 10\,$kpc) is much smaller than the coherence length of the 
relative velocity field associated with acoustic oscillations (few Mpc).  This will be critical for our use of moving-background perturbation theory to follow
early structure formation.

\subsection{Fluid equations}

After recombination the small-scale inhomogeneities in the photons and neutrinos are rapidly washed out by free-streaming, and the dark energy is not yet dynamically 
significant.  Also on small scales we can ignore the general relativistic (higher order in $aH/k$) terms.
Thus we can write the evolution equations as the pressureless Navier-Stokes equations for the CDM, the Navier-Stokes equations including pressure for 
the baryons, and the Poisson equation for the gravitational potential (e.g. \cite{2002PhR...367....1B}):
\beqa
\frac{\partial \delta_{\rm c}}{\partial t} + a^{-1}{\bf v}_{\rm c}\cdot \nabla \delta_{\rm c} &=& - a^{-1}(1+\delta_{\rm c})\nabla\cdot{\bf v}_{\rm c},
\nonumber \\
\frac{\partial {\bf v}_{\rm c}}{\partial t} + a^{-1}({\bf v}_{\rm c}\cdot \nabla){\bf v}_{\rm c} &=& -\frac{\nabla \Phi}{a} - H{\bf v}_{\rm c},
\nonumber \\
\frac{\partial \delta_{\rm b}}{\partial t} + a^{-1}{\bf v}_{\rm b}\cdot \nabla \delta_{\rm b} &=& - a^{-1}(1+\delta_{\rm b})\nabla\cdot{\bf v}_{\rm b},
\nonumber \\
\frac{\partial {\bf v}_{\rm b}}{\partial t} + a^{-1}({\bf v}_{\rm b}\cdot \nabla){\bf v}_{\rm b} &=& -\frac{\nabla \Phi}{a} - H{\bf v}_{\rm b}
\nonumber \\
&& - a^{-1}c_{\rm s}^2\nabla\delta_{\rm b},
{\rm and}
\nonumber \\
a^{-2}\nabla^2\Phi &=& 4\pi G \bar{\rho}_{\rm m} \delta_{\rm m}.
\label{eq:pert}
\eeqa
Here $\Phi$ is total gravitational potential and subscripts c, b, and m stand for dark matter, baryons and total matter respectively.

A more complete treatment would also follow the baryon entropy \cite{2005MNRAS.362.1047N} and ionization fraction 
\cite{2007PhRvD..76f3001L} perturbations.  We have not done this here, but we note that the moving background 
perturbation theory approach described here could be extended to accommodate these additional variables.

The standard way to do large scale structure perturbation theory is to Taylor-expand in powers of the primordial 
perturbations, e.g. $\delta_{\rm c} = \delta_{\rm c}^{(1)} + \delta_{\rm c}^{(2)} + \delta_{\rm c}^{(3)} + ...$.  One 
may then use the linear terms in the above equation to describe the behavior at
order $n$; for $n\ge 2$, the quadratic terms in Eq.~(\ref{eq:pert}) may be treated as a source for the order-$n$ 
perturbation, written hierarchically in terms of orders $<n$ \cite{1984ApJ...279..499F, 1986ApJ...311....6G, 
1994ApJ...431..495J, 1994MNRAS.267.1020P}.  This approach can even be extended to include both baryons and CDM 
\cite{2009ApJ...700..705S}.
In our case, this is not appropriate: since there are relative bulk flow 
velocities between the baryons and CDM with Mach numbers of order 10, it follows that for perturbations at the 
baryonic Jeans scale the baryon and CDM components will be advected relative to each other by up to several 
perturbation wavelengths.  Whether the standard perturbation series will converge in this case is an open question; 
even if it does, we expect that many orders in perturbation theory would be required.  Therefore, we desire an 
alternative method to follow the growth of the smallest structures.

\subsection{Moving-background perturbation theory (MBPT)}

Our preferred method of following the earliest structures is to do a perturbation analysis on a background where the baryons move relative to the CDM.  The idea is 
that in the absence of density perturbations, but in the presence of a bulk relative velocity, there exists an exact solution to Eq.~(\ref{eq:pert}):
\begin{eqnarray}
{\bf v}_{\rm c}({\bf x},t) &=& {\bf v}_{\rm c}^{\rm(bg)}(t),
\nonumber \\
{\bf v}_{\rm b}({\bf x},t) &=& {\bf v}_{\rm b}^{\rm(bg)}(t),
{\rm ~~and}
\nonumber \\
\Phi&=&\delta_{\rm c}=\delta_{\rm b}=0,
\label{eq:exact}
\end{eqnarray}
where ${\bf v}_{\rm c}^{\rm(bg)}$ and ${\bf v}_{\rm b}^{\rm(bg)}$ are constant with position and have temporal dependence $\propto 1/a(t)$.  Without loss of 
generality, one may boost to a different reference frame and set one of these (e.g. ${\bf v}_{\rm b}^{\rm(bg)}$) equal to zero.

Since the relative velocity of the baryons and CDM is coherent over scales of several comoving Mpc, whereas the scales of direct interest for us are at a few 
baryonic Jeans lengths ($\sim 10$ comoving kpc), the above moving background is an appropriate zeroeth-order solution in any small ($\sim 1$ Mpc) region of the 
Universe.  Thus we can imagine the Universe as composed of many individual patches, each of which has a different relative velocity ${\bf v}_{\rm bc}^{\rm(bg)}$.  
Small fluctuations on this background grow due to gravitational instability; their early stages of collapse can be modeled using linear perturbation theory around 
Eq.~(\ref{eq:exact}) using the local value of ${\bf v}_{\rm bc}^{\rm(bg)}$.

Perturbing around Eq.~(\ref{eq:exact}), and writing the perturbation variables ${\bf u}_{\rm b,c}$,
\begin{equation}
{\bf v}_{\rm b}({\bf x},t) = {\bf v}_{\rm b}^{\rm(bg)}(t) + {\bf u}_{\rm b}({\bf x},t)
\end{equation}
and similarly for ${\bf u}_{\rm c}$, we may transform Eq.~(\ref{eq:pert}) into a system of equations involving ${\bf u}_{\rm b,c}$ instead of ${\bf v}_{\rm b,c}$.  
Working only to first order in the new perturbation variables $\{\delta_{\rm c},{\bf u}_{\rm c},\delta_{\rm b},{\bf u}_{\rm b},\Phi\}$ we find:
\beqa
\frac{\partial \delta_{\rm c}}{\partial t} + a^{-1}{\bf v}^{\rm(bg)}_{\rm c}\cdot \nabla \delta_{\rm c} &=& - a^{-1}\nabla\cdot{\bf u}_{\rm c},
\nonumber \\  
\frac{\partial {\bf u}_{\rm c}}{\partial t} + a^{-1}({\bf v}_{\rm c}^{\rm(bg)}\cdot \nabla){\bf u}_{\rm c} &=& -\frac{\nabla \Phi}{a} - H{\bf u}_{\rm c},
\nonumber \\
\frac{\partial \delta_{\rm b}}{\partial t} + a^{-1}{\bf v}_{\rm b}^{\rm(bg)}\cdot \nabla \delta_{\rm b} &=& - a^{-1}\nabla\cdot{\bf u}_{\rm b},
\nonumber \\
\frac{\partial {\bf u}_{\rm b}}{\partial t} + a^{-1}({\bf v}_{\rm b}^{\rm(bg)}\cdot \nabla){\bf u}_{\rm b} &=& -\frac{\nabla \Phi}{a} - H{\bf u}_{\rm b}
\nonumber \\
&& - ac_{\rm s}^2\nabla\delta_{\rm b},
{\rm ~and}
\nonumber \\
a^{-2}\nabla^2\Phi &=& 4\pi G \bar{\rho}_{\rm m} \delta_{\rm m}.
\label{eq:MBPT}
\eeqa

It is convenient to 
transform these equations into Fourier space and use the last equation to eliminate $\Phi$.  We may also re-write the velocity equations in terms of the divergence 
$\theta$, with ${\bf u}_{\rm i}({\bf k}) = -iak^{-2}{\bf k}\theta_{\rm i}({\bf k})$ (i=b or c),
since under the approximation of barotropic flow of the baryons the vorticity remains zero until the development of structure formation shocks.
We may also work in the bulk baryon frame, i.e. we may set
\beq
{\bf v}_{\rm b}^{\rm(bg)} = 0 {\rm ~~and~~}
{\bf v}_{\rm c}^{\rm(bg)} = -{\bf v}_{\rm bc}^{\rm(bg)}(t).
\eeq
This reduces our system of equations to
\beqa
\frac{\partial \delta_{\rm c}}{\partial t} &=&
\frac ia{\bf v}^{\rm(bg)}_{\rm bc}\cdot{\bf k} \delta_{\rm c}
 - \theta_{\rm c},
\nonumber \\
\frac{\partial \theta_{\rm c}}{\partial t} &=&
\frac ia{\bf v}_{\rm bc}^{\rm(bg)}\cdot{\bf k}\theta_{\rm c}
 -\frac{3H^2}{2}
(\Omega_{\rm c}\delta_{\rm c} + \Omega_{\rm b}\delta_{\rm b}) 
- 2H\theta_{\rm c},
\nonumber \\
\frac{\partial \delta_{\rm b}}{\partial t} &=& -\theta_{\rm b}, {\rm ~and}
\nonumber \\
\frac{\partial \theta_{\rm b}}{\partial t} &=& -\frac{3H^2}{2}
(\Omega_{\rm c}\delta_{\rm c} + \Omega_{\rm b}\delta_{\rm b}) 
- 2H\theta_{\rm b}
 + \frac{c_{\rm s}^2k^2}{a^2} \delta_{\rm b}.
\label{eq:evoleqn}
\eeqa
Note that $\Omega_{\rm c,b}$ are evaluated at the appropriate redshift rather than taking on their present-day values.
Our code evolves these equations, albeit with the scale factor $a$ as the independent variable, which can be accomplished using the replacement $\partial/\partial t = 
aH\,\partial/\partial a$.  It is important to note the time dependence ${\bf v}_{\rm bc}^{\rm(bg)}\propto 1/a(t)$ when evolving Eq.~(\ref{eq:evoleqn}).


On large scales $k\ll k_{\rm vbc}\sim 40\,$Mpc$^{-1}$ used for galaxy clustering and even for Lyman-$\alpha$ forest studies, the ${\bf v}_{\rm bc}$ terms in 
Eq.~(\ref{eq:evoleqn}) are negligible.  However, at $k\gtrsim k_{\rm vbc}$, the advection terms become comparable to or larger than the Hubble expansion rate, and 
they must be taken into account.  Note that this is true even if one's interest is only in the CDM perturbations, since the baryons contribute 17\%\ of the energy 
density and hence their perturbations are important in Eq.~(\ref{eq:evoleqn}).  (As an extreme example, below the Jeans scale $k>k_{\rm J}$, the growth of structure 
in the CDM switches from the ``standard'' $\delta\propto a$ growth to a slower growth $\delta\propto a^{\sqrt{25-24\Omega_{\rm b}}/4-1/4}$ 
\cite{1998ApJ...498..497H, 2005MNRAS.362.1047N}.)

\subsection{Small-scale transfer function and matter power spectrum}

The usual way to describe the small-scale distribution of matter is to derive a transfer function $T(k)$ that maps primordial to final potentials, and then to write the matter power spectrum $P_{\rm 
m}(k)$, equal to the primordial power spectrum times $|T(k)|^2$ (times normalization factors \cite{2003moco.book.....D}).  We may solve the transfer functions including the relative velocity effect by 
solving
the system of equations, Eq.~(\ref{eq:evoleqn}).  We evolve these from the redshift of recombination, where initial conditions are 
determined using {\sc Cmbfast} \cite{1996ApJ...469..437S}, to $z=40$.  The resulting transfer function, evaluated at $z=40$, is clearly a function of the local relative velocity ${\bf v}_{\rm 
bc}^{\rm(bg)}$ and also of the angle $\vartheta$ 
between the direction of the wave vector ${\bf k}$ and ${\bf v}_{\rm bc}^{\rm(bg)}$.

We may determine a local {\em isotropically averaged} power spectrum $P_{\rm loc,m}(k;v_{\rm bc})$ by averaging over the direction of ${\bf k}$, i.e. we write:
\begin{equation}
P_{\rm loc,m}(k;v_{\rm bc}) = P_\zeta(k) \,\frac12\int_0^\pi
\left| \frac{\delta_{\rm m}({\bf k};{\bf v}_{\rm bc})}{\zeta({\bf k})} \right|^2
\,\sin\vartheta\, d\vartheta,
\label{eq:Ploc}
\end{equation}
where $\zeta({\bf k})$ is the primordial curvature perturbation and $P_\zeta(k)$ is its power spectrum.  This power spectrum depends only on the magnitude of $v_{\rm bc}$ (we drop the superscript 
$^{\rm(bg)}$ to reduce clutter).
In order to determine an 
overall effect on the small-scale matter power spectrum we need to average over a large number of coherence regions with different $v_{\rm bc}$.  The latter arises from linear perturbation theory on 
large scales and hence is well-described by a 3-dimensional Gaussian distribution with variance per axis
\beq
\sigma^2_{\rm vbc} = \frac13\left\langle \left| {\bf v}_{\rm bc}({\bf x}) \right|^2\right\rangle.
\eeq
We may then average Eq.~(\ref{eq:Ploc}) to obtain a globally averaged matter power spectrum.

\begin{figure}
\includegraphics[width=3.4in]{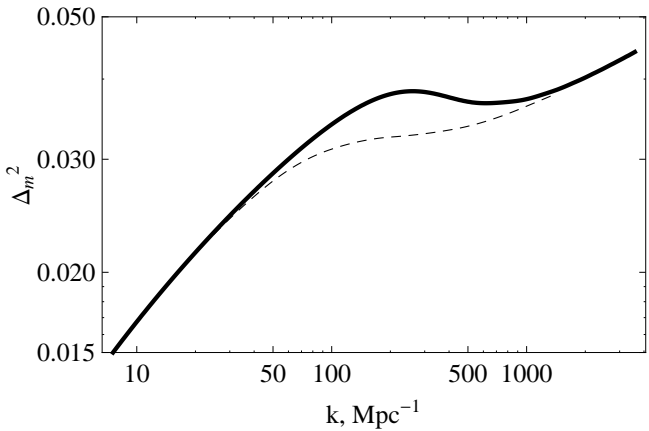}
\caption{\label{fig:PkV} Power spectrum of matter distribution in the first order CDM model (solid line) and with the $v_{\rm bc}$ effect included (dashed line) at the redshift of $z = 40$.}
\end{figure}

Intuitively, we expect the relative velocity effect to suppress the small-scale power spectrum, since the moving baryons have pressure $\sim \rho_{\rm b}v_{\rm bc}^2$ in the CDM frame.
This suppression is shown in Figure~\ref{fig:PkV} where we plot $\Delta_{\rm m}^2(k)\equiv [k^3/(2\pi^2)]P_{\rm m}(k)$ for the cases with and without the effect 
of relative velocity. The power is most strongly suppressed around the Jeans scale $k_{\rm J} = aH/c_{\rm s} \sim 200 \ $Mpc$^{-1}$, where a difference of $\sim 15$\%\ is computed.

The effect of $v_{bc}$ is not limited to the suppression of power on small scales, but rather has an important implication for the distribution of the first bound structures with respect to matter 
distribution as well as for the number densities of the first halos. To study these effects we ran a set of simulations in which the large-scale density and velocity fields were generated according to 
linear perturbation theory.  We then used analytical (Press-Schechter) arguments to predict the number of haloes formed in each cell of our cosmological box.  This hybrid approach is computationally feasible on a single desktop computer since it does not have to numerically follow the small-scale modes, and should capture the rough magnitude of the effect.  However, ultimately a simulation that follows the full nonlinear evolution of the small-scale modes will be required. The key reason for using approximate methods in the present study, as opposed to a full hydrodynamic numerical simulation, is our desire to introduce the concept of relative velocity effect in the simplest and most intuitive way while allowing more detailed study to be performed by other research groups in an unbiased manner. 

\section{The abundance and clustering of early haloes}
\label{sec:sim}

We now investigate the formation of the first baryonic objects, taking account of the relative velocity effect.  This is a difficult problem, which we only partially solve in this paper: one has acoustic 
oscillations in the photon-baryon plasma that travel $\sim 140$ Mpc, and simultaneously one must resolve the baryon Jeans scale.  We provide a computation based on the formalism described above: we 
generate a
realization of a Gaussian random primordial perturbation on a 3D grid, and then to each cell we assign an overdensity $\delta_{\rm l}$ (where the ``l'' referes to long-wavelength modes) using periodic boundary condition and a relative 
velocity $\bf{v_{\rm bc}}$ derived from the linear density field. Initial values of $\delta_{\rm l}$ are obtained using the linear perturbation theory, as there is no significant difference between the theory with and without relative motion effect before the time of recombination when the values of $\delta_{\rm l}$ are formed.  Then, within each cell, we use the peak-background split to compute the number density of haloes.  The new twist is that the small-scale power spectrum is modulated by the 
large-scale $v_{\rm bc}$.  (In some ways, this is similar to the modification of the peak-background split used for local $f_{\rm NL}$-type non-Gaussianity studies 
\cite{2008PhRvD..77l3514D, 2008JCAP...08..031S}, except that in our case the modulation of the small-scale power spectrum is a 
result of the advection process and arises even in standard $\Lambda$CDM cosmology with Gaussian adiabatic initial conditions.)  This of course depends on an analytic model for the halo mass function; we 
have used the Press-Schechter model \cite{1974ApJ...187..425P, 1991ApJ...379..440B}.  The validity of Press-Schechter for any precise calculation
is dubious -- particularly since it is being applied here with an anisotropic local power 
spectrum -- but we expect that the qualitative results (a scale-dependent enhancement in the bias and stochasticity at large scales, with acoustic oscillations in each) would still arise in a more 
accurate treatment.

\subsection{Peak-background split}

The collapse of the first halos can be conveniently treated in the framework of the peak-background split formalism~\cite{1989MNRAS.237.1127C}, in which the growth of small-scale inhomogeneities depends 
on the large scale overdensity. One can split the density field into a long-wavelength piece $\delta_{\rm l}$ and a
short-wavelength piece $\delta_{\rm s}$:
\begin{equation}
  \rho(\vec{x}) = \bar{\rho} \left[1+\delta_{\rm l}({\bf x})+\delta_{\rm s}({\bf x}) \right].
\end{equation}
In any region, the number density of haloes of any given type generally depends on the large-scale overdensity $\delta_{\rm l}$, and on the statistics of the small-scale perturbations $\delta_{\rm s}$ 
(in particular, their local power spectrum).  In the usual case where the small and large-scale perturbations are independent, the number density becomes purely a function of the large-scale overdensity 
plus a stochastic component $\epsilon$ with $\langle\epsilon({\bf x})\rangle=0$; Taylor-expanding in $\delta_{\rm l}$ gives
\begin{equation}
  n(\vec{x}) = \bar{n} \left[1+b_0\delta_l({\bf x}) \right] + \epsilon({\bf x}).
\label{eq:nx}
\end{equation}
The bias is then
\begin{equation}
b_0 = \bar{n}^{-1} \frac{\partial n}{\partial \delta_{\rm l}}.
\label{bl}
\end{equation}
This argument leads to a generically scale-independent bias at sufficiently large scales (with the addition of a Poisson or halo-shot-noise term \cite{2000MNRAS.318..203S, 2006PhRvD..74j3512M}).

When the relative motion of dark matter and baryons is introduced the growth of small scale overdensities becomes dependent on the local value of the relative velocity.  Equation~(\ref{eq:nx}) then 
generalizes to
\beq
n(\vec{x}) = n\left[ \delta_{\rm l}({\bf x}), v_{\rm bc}({\bf x})\right] + \epsilon({\bf x}).
\label{eq:nx2}
\eeq
At Mach numbers of order 10, it is not clear whether we can Taylor-expand in $v_{\rm bc}$.  Therefore, our strategy will be to re-compute $n({\bf x})$ 
in each cell, using the Press-Schechter 
conditional mass function, i.e. the number of haloes per unit comoving volume per $\ln M$:
\beqa
N(M|\delta_{\rm l},v_{\rm bc}) &=& \frac{\bar{\rho}}{\sqrt{2\pi}}\frac{\delta_{\rm sc} - \delta_{\rm l}}{\sigma^2}\left|\frac{d\sigma}{dM}\right|
\nonumber \\ && \times (1+\delta_{\rm l})
\exp[-\frac{(\delta_{\rm sc} - \delta_{\rm l})^2}{2\sigma^2}],
\label{eq:PS}
\eeqa
where $\delta_{\rm sc}$ is critical overdensity of spherical collapse, $M$ is the halo mass and $N(M|\delta_{\rm l},v_{\rm bc})$ has units of Mpc$^{-3}$.The factor $1+\delta_{\rm l}$ is the conversion from the Lagrangian volume element (in which the Press-Schechter formalism is native) to the Eulerian volume element.  Here $\sigma^2$ is the variance of the density field smoothed with the top-hat window function,
\beq
\sigma^2(M,\vec{v}_{\rm bc}) = \int \Delta_m^2(k,\vec{v}_{\rm bc}) |W(k,R)|^2 \frac{dk}{k},
\eeq
where for $\Delta_m^2(k,\vec{v}_{\rm bc})$ we use the isotropically averaged local matter power spectrum.  In principle, one should follow here the 
formation of haloes in a statistically anisotropic density field.  This will ultimately require a hydrodynamic (or at least $N$-body) simulation to 
achieve results that can be used for detailed analysis.  
However, for the moment we use the Press-Schechter formalism;
the top-hat window function in Fourier space can be written as $W(k,R)=3j_1(kR)/(kR)$,
where the smoothing scale $R$ is determined by the halo mass $M = \frac43\pi R^3$.

In our case -- unlike the usual case of purely Gaussian density perturbations -- 
$\sigma^2(M)$ and hence $d\sigma/dM$ are explicit functions of relative velocity and hence will change from place 
to place. 

\subsection{Simulation parameters}

Our fiducial box size is $ 1365^3\,$Mpc$^3$.  The box is divided into smaller boxes each of the size of coherence length for the relative velocity field, $\Delta = 3$ Mpc.  Initial (i.e. at $z_{\rm 
dec}$) 
density
and velocity distributions are generated using the {\sc Cmbfast} power spectrum computations \cite{1996ApJ...469..437S} and smoothed with a Gaussian 
window function with scale length $k_{\rm scale}=\pi/(\sqrt3\,\Delta)$.
The 
smoothing is necessary to avoid aliasing and spurious effects from the finite resolution of the simulation. For each small box we generate initial values of ${\bf v}_{\rm bc}$ and large scale overdensity
$\delta_{\rm l}$ at the time of recombination using {\sc Cmbfast}. Halo number densities in each cell are then inferred from Eq.~(\ref{eq:PS}).

\begin{figure}
\includegraphics[width=3.4in]{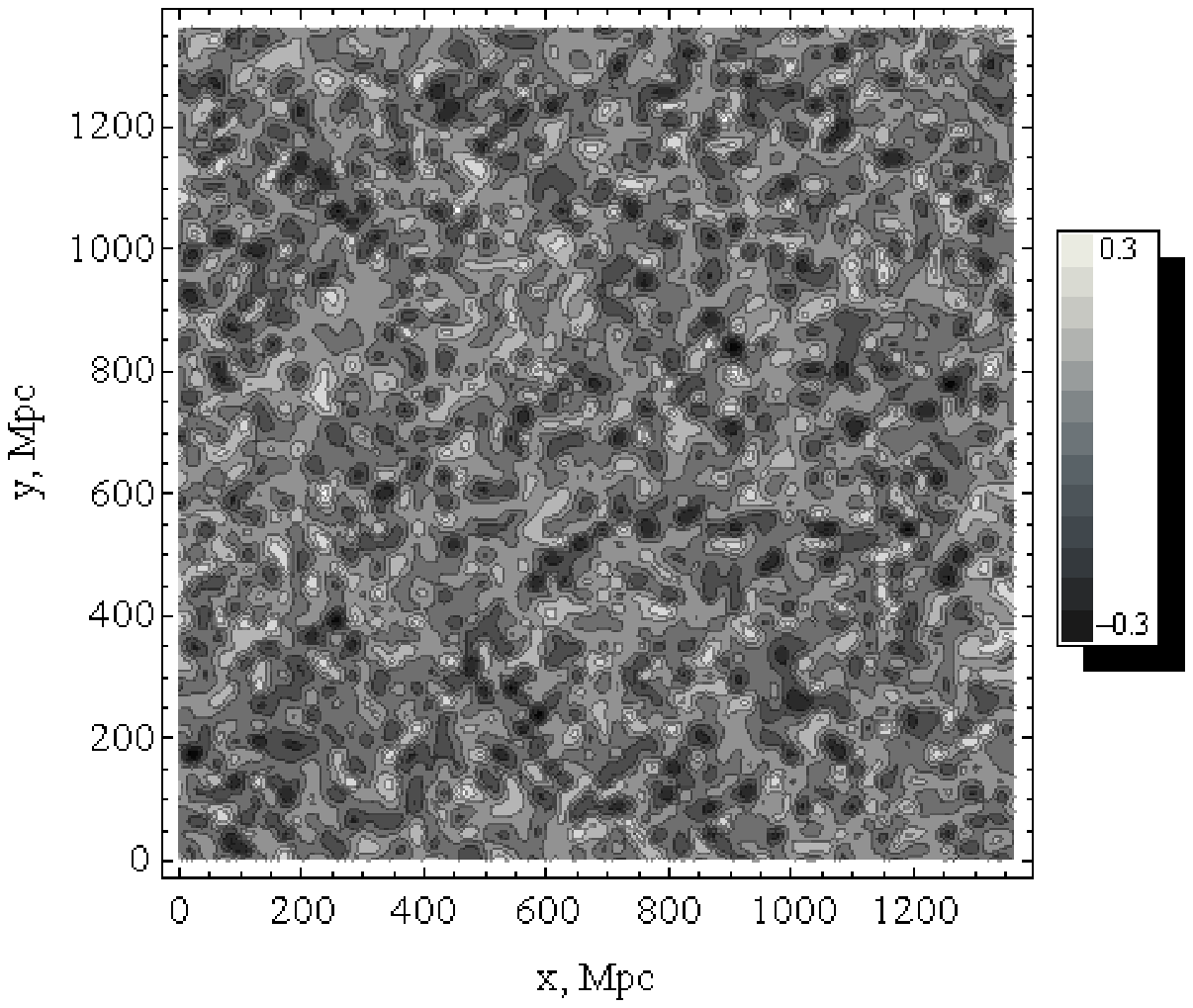}
\includegraphics[width=3.4in]{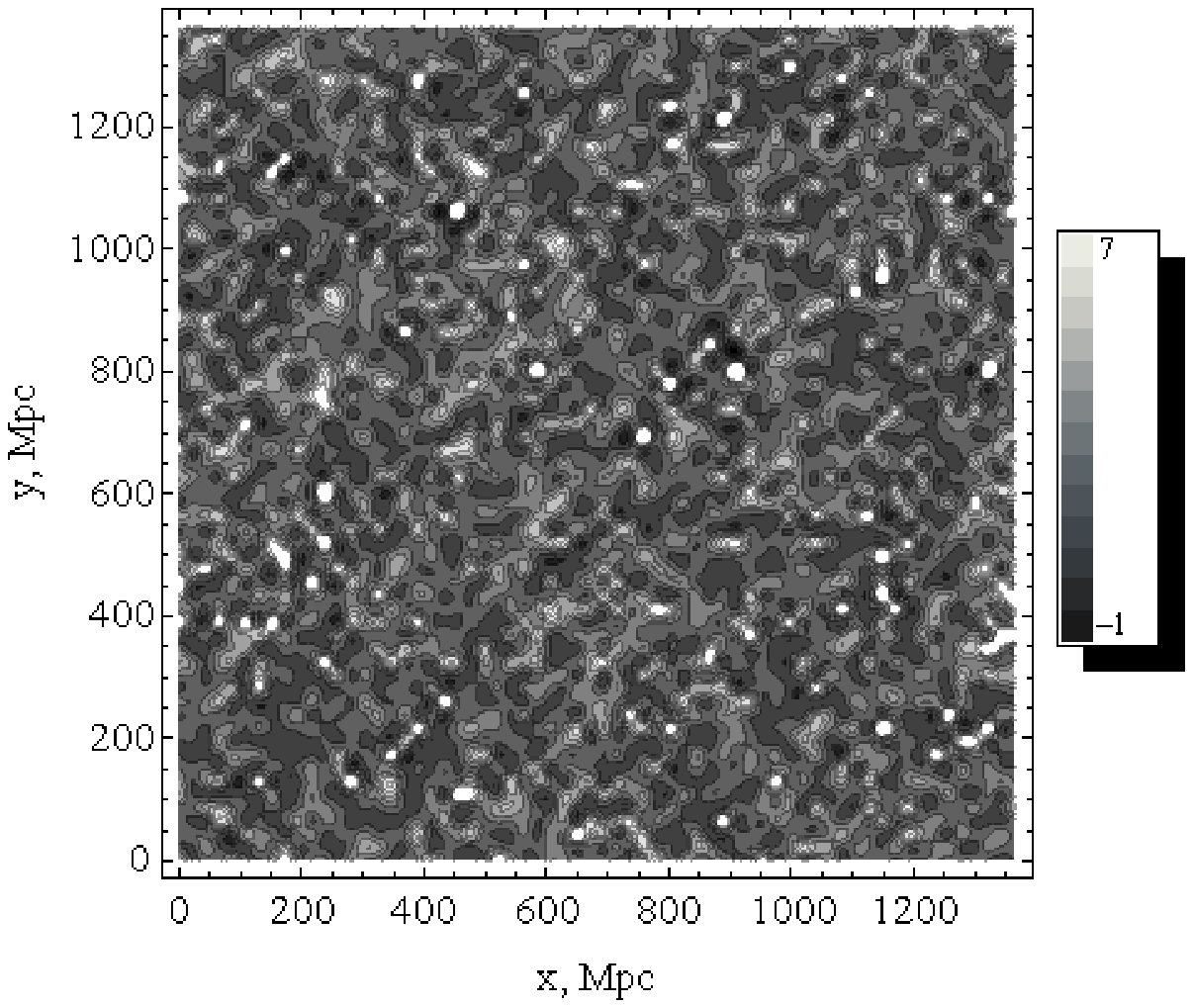}
\includegraphics[width=3.4in]{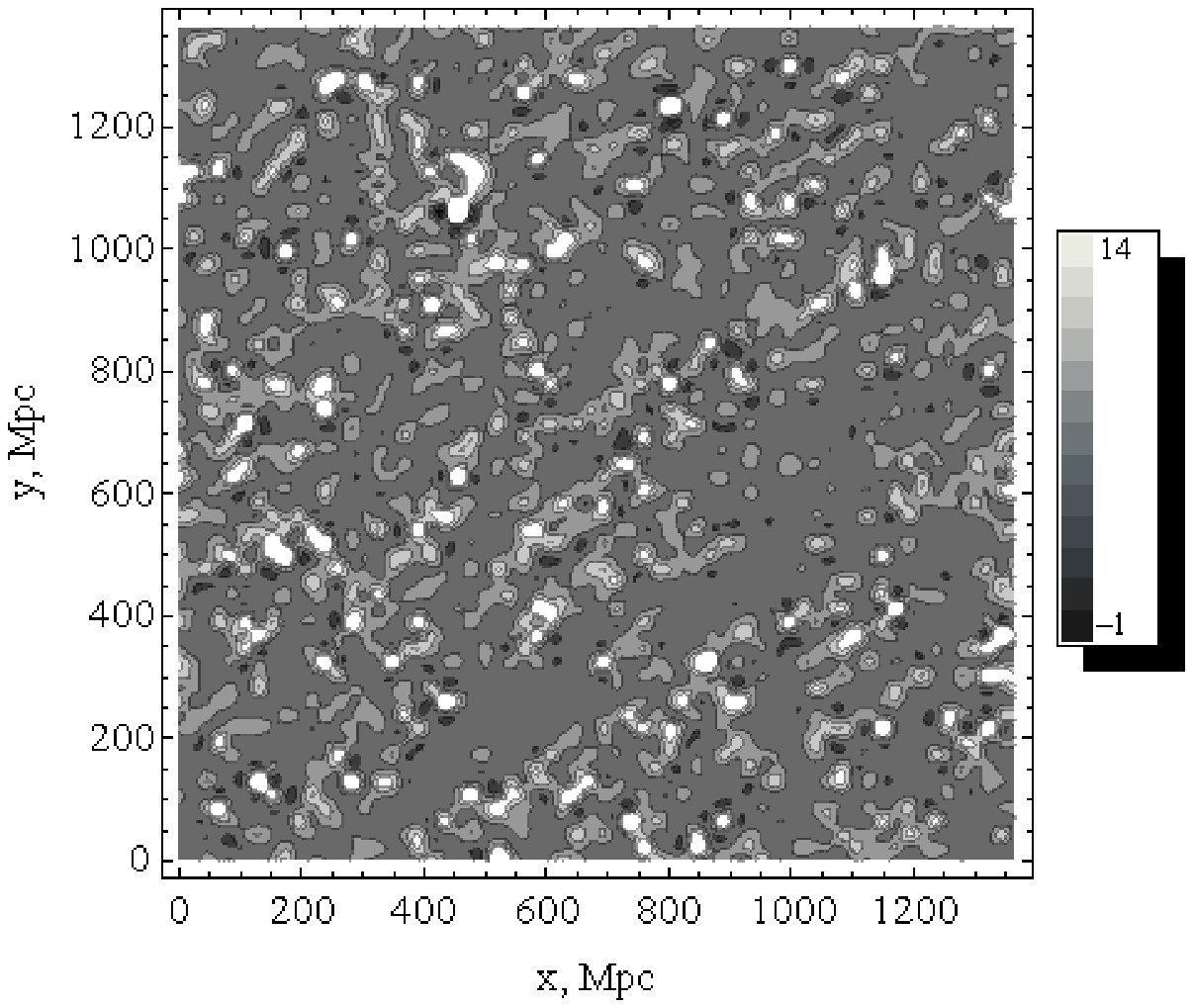}

\caption{\label{fig:Density}{\em Top panel}: The matter density contrast $\delta_m$ on a 2D slice of the 3D simulation box. The halo density contrast $\delta_n$ for $M_{\rm halo}=10^6\,M_\odot$ on the same slice with $V_{bc} = 0$ (Middle panel) and with $V_{bc} \neq 0$ (Bottom panel).  All panels are at $z = 40$.
}
\end{figure}

In Figure~\ref{fig:Density}, we show an example output from this procedure.  The top panel shows the matter density contrast and the bottom two panels the halo density contrast for $M_{\rm halo}=10^6\,M_\odot$ without (Middle panel) and with (Bottom panel) the relative motion effect at the redshift of $z = 40$.  Note that the structures in matter and halo overdensities, while correlated, are not identical. Comparison of the halo density contrasts for two different cases clearly shows the importance of relative velocity effect on the formation of first bound objects. 

\subsection{Halo abundance}

\begin{figure}
\includegraphics[width=3.4in]{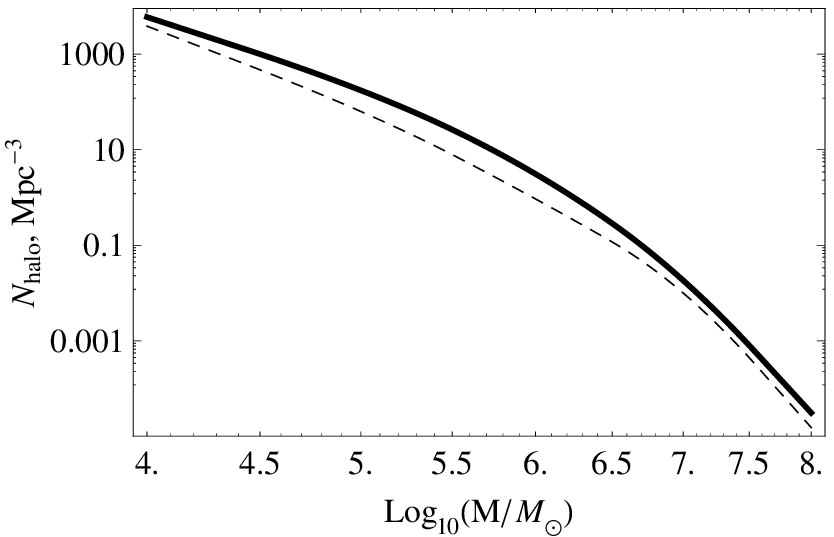}
\includegraphics[width=3.4in]{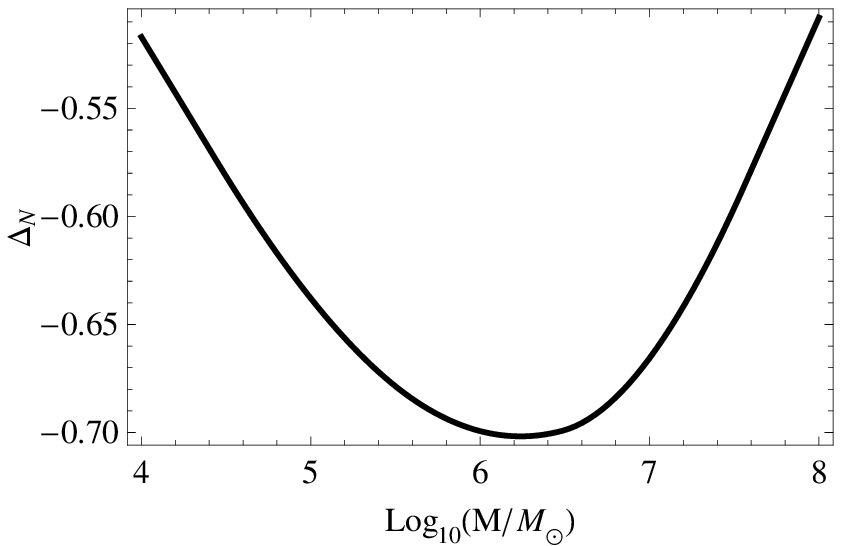}
\caption{\label{fig:Nav}{\em Top panel}: The number density of dark matter haloes produced in our simulation box without the effect of relative velocity (solid line) and with the effect (dashed line). {\em Bottom panel}: The relative decrease in the number density of haloes as a function of the halo mass. Number densities in our simulation correspond to the redshift of $z = 40$.} 
\end{figure}

To illustrate the effect of relative velocity  on the abundance of small haloes, we calculate 
number densities of collapsed halos with and without relative velocity. The decrease in number density is quantified by 
\beq
\Delta_N = \frac{\bar{N}_{\rm vbc} - \bar{N}_{0}}{\bar{N}_{0}},
\label{eq:dneq}
\eeq
where $\bar{N}_{0}$ and $\bar{N}_{vbc}$ are average number densities of halos without the effect of $\vec{v_{bc}}$ and with it. The comparison of the two cases shows that the number density of haloes is suppressed 
by more than 60\% at the mass scale of $M \sim 10^6 M_{\odot}$, as can be seen in Figure~\ref{fig:Nav}. Note that the strongest suppression occurs for 
halo masses of $10^{6.3}\,M_\odot$, corresponding to top-hat scales of 20 kpc comoving, i.e. near $k_{\rm vbc}^{-1}$. We emphasize that 
the results provided in the figure are based on the Press-Schecter formalism and are a good qualitative guide, but should not be interpreted 
quantitatively.

\subsection{Bias, stochasticity and the large scale distribution of early haloes}

The introduction of relative motions modifies the correlation between the first halos and the matter distributions rendering bias parameter scale dependent. Because of the non-linear terms in the 
evolution 
equations dark matter and baryonic matter evolve out of phase and the growth of the overdense regions become dependent on both $\delta_{\rm l}$ and $v_{\rm bc}$.
 
To quantify this effect we calculated halo overdensity using number densities of halos in each of the small boxes from our simulation:
\beq
\delta_n(M,x) = \frac{N(M,x) - \bar{N}(M)}{\bar{N}(M)}.
\eeq
Next, we calculate power spectra of halos of various masses:
\beq
(2\pi)^3\delta_{\rm D}({\bf k}-{\bf k}')P_{\rm hh}(k|M) = \langle \delta_n(M,{\bf k}) \delta_n(M,{\bf k}')^\ast\rangle,
\eeq
where $\delta_{\rm D}$ is the Dirac $\delta$-function.

The difference between this case and the case neglecting $v_{\rm bc}$ can be illustrated by defining a bias correction parameter $\Delta b(k)$:
\beq
P_{\rm hh}(k) = b_0^2\left[1+\frac{\Delta b(k)}{b_0}\right]^2 P_{\rm mm}(k),
\eeq
where $b_0$ is a Gaussian scale independent bias, which in the Press-Schechter formalism is given by:
\beq
b_0 = \frac{\delta_{\rm sc}}{\sigma^2} - \frac{1}{\delta_{\rm sc}} +1.
\eeq

\begin{figure}
\includegraphics[width=3.4in]{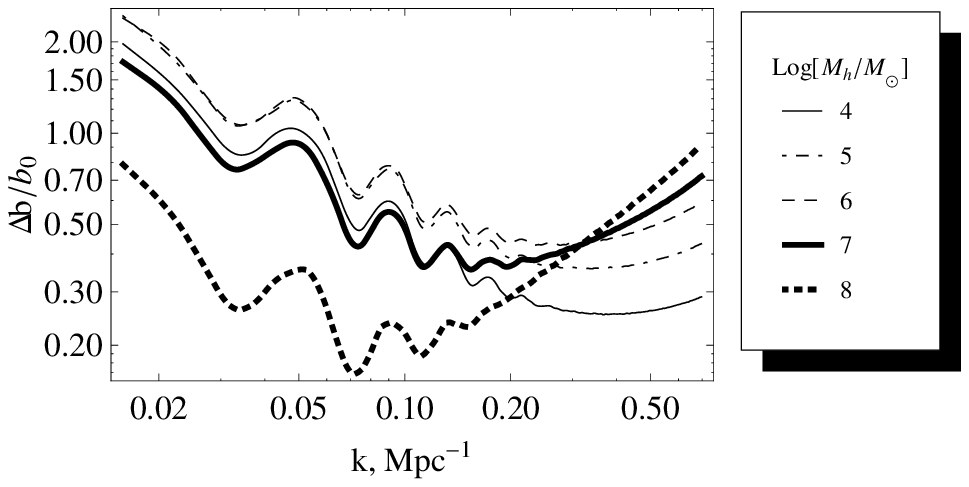}
\includegraphics[width=3.4in]{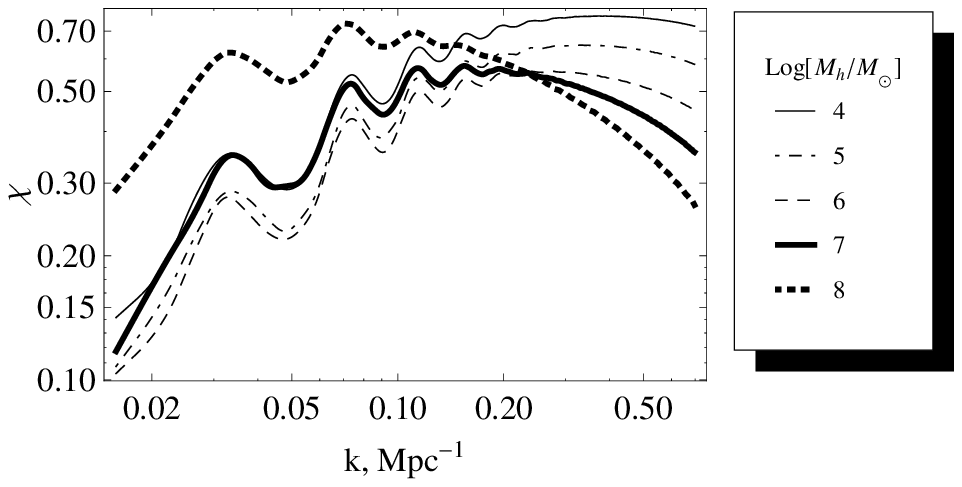}
\caption{\label{fig:db}The correction to the bias parameter $\Delta b$ (top panel) and the stochasticity $\chi=r_{\rm hm}^2$ (bottom panel) for various halo masses at $z=40$.  The
solid curve corresponds to $M_h = 10^4 M_{\odot}$; the thick-solid to $M_h = 10^5 M_{\odot}$; the dashed to $M_h = 10^6 
M_{\odot}$; the dot-dashed to $M_h = 10^7 M_{\odot}$; and the dotted to $M_h = 10^8 M_{\odot}$. In the first order CDM model $\Delta b  = 0$ and $\chi=1$ on large scales.
The enhancement of bias on small scales $k>0.3\,$Mpc$^{-1}$ is due to the nonlinear dependence of halo abundance on $\delta_{\rm l}$.}
\end{figure}

Using these results along with the matter power spectrum we can obtain the scale-dependent component of the bias parameter $\Delta b$ which is plotted in the top panel of Figure~\ref{fig:db} for various 
halo masses. The plot 
shows that for halos with mass $M \sim 10^4$--$10^8 M_{\odot}$ there is a significant increase of the bias.
The effect of ${\bf v}_{\rm bc}$ becomes less important for heavier and lighter masses which can be expected from the analysis of power suppression in Fig.~\ref{fig:PkV}.  This is 
principally a consequence of the fact that for very massive haloes the baryons advect through a distance that is only a small fraction of the halo scale $R$, and hence this advection does not affect the 
formation of the halo; whereas for the lowest-mass haloes, whose scale $R$ is smaller than the baryon Jeans length, the baryons can be treated as homogeneous irrespective of their velocity.

To further understand the importance of $v_{\rm bc}$ we calculate the stochasticity of the halos relative to the matter. In the bottom panel of Figure~\ref{fig:db} we plot the stochasticity $\chi$ as a 
function of 
wave number 
$k$ for various halo masses.  The stochasticity is defined as:
\beq
\chi = \frac{P_{\rm hm}^2(k)}{P_{\rm hh}(k)P_{\rm mm}(k)},
\eeq
where the cross power spectrum $P_{\rm hm}$ is defined
via $(2\pi)^3\delta_{\rm D}({\bf k}-{\bf k}')P_{\rm hm}(k) = \langle \delta_{\rm h}(M,{\bf k}) \delta_{\rm m}^\ast({\bf 
k}')\rangle$. In the linear 
theory, without consideration of the $v_{\rm bc}$ effect one would have $\chi=1$ (modulo Poisson corrections as described above).

We checked the convergence of our results by running the simulation with varying box sizes and varying $\Delta$. Specifically we tried runs with $\Delta=4$ Mpc, and found changes of less than $1\%$ in 
the stochasticity and bias over the range $0.2<k<1\,$Mpc$^{-1}$ at $M_{\rm halo}=10^4M_\odot$, whereas using $\Delta=6$ Mpc produces change greater than $5\%$ and distorts functional forms of both bias 
and stochastisity at $k>0.1$ Mpc$^{-1}$. Similarly, increasing the box size to $2275^3$ Mpc$^3$, whith fixed $\Delta$ did not produce observable change in $\chi$ and $\Delta b$, whereas decreasing the 
box size to $1000^3$ Mpc$^3$ changes our results by $\sim 5\%$ at $k < 0.1\,$Mpc$^{-1}$. As a test, we repeated the analysis setting $v_{bc} = 0$ to recover the `standard' picture with a scale-independent halo bias and the stochastisity consistent with the linear theory prediction. Specifically, we found that at $k < 0.2 \ Mpc^{-1}$ the stochastisity is  $0.98 < \chi < 1$. The small deviation from unity can be explained by the fact that mapping from the overdensity $\delta_{\rm l}$ to the number density of halos $N(M|\delta_{\rm l},v_{\rm bc})$ is not exactly linear even in Press-Schecter model. 

\begin{figure}
\includegraphics[width=3.4in]{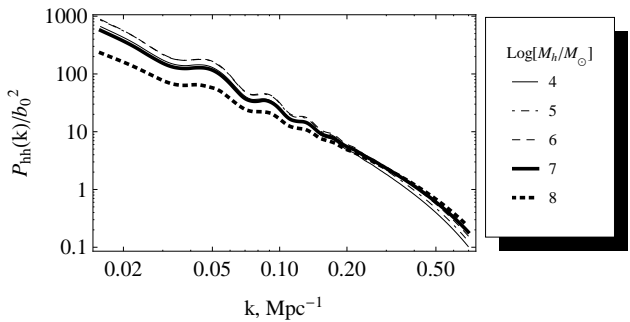}

\caption{\label{fig:phh}Scaled halo-halo power spectrum $P_{hh}(k)/b_0^2(M_h)$ at the redshift of $z = 40$ for various halo masses. The
solid curve corresponds to $M_h = 10^4 M_{\odot}$; the thick-solid to $M_h = 10^5 M_{\odot}$; the dashed to $M_h = 10^6 
M_{\odot}$; the dot-dashed to $M_h = 10^7 M_{\odot}$; and the dotted to $M_h = 10^8 M_{\odot}$.}
\end{figure}

We also would like to mention that Figure~\ref{fig:db} exhibits strong oscillations of $\Delta b$ which correspond to the BAO in the matter power 
spectrum. This means that the signal of the BAO in the halo power spectrum for these halo masses is very different from that of the dark matter. To 
illustrate this point we plot the actual scaled halo power spectrum $P_{hh}(k)/b_0^2(M_h)$ (Figure~\ref{fig:phh}) for different halo masses covering the range from $M_h = 10^4$ to $M_h = 10^8$ that shows the behavior of the BAO signal. Although these are very low-mass haloes compared to those probed by BAO surveys ($M>10^{11}M_\odot$), they are the seeds of present day galaxies, and their subsequent evolution might alter the BAO signal in the galaxy power spectrum at lower redshifts. As with other interesting applications of the relative velocity effect we relegate detailed analysis of the this problem to a future study.

\section{Conclusions and prospects}
\label{sec:conc}

We have shown that the relative velocity of baryonic and dark matter fluids plays an important role in the formation and evolution of small-scale 
structure of the early universe.  In light of the increasing interest in reionization, high-redshift galaxy clustering, and dark matter substructure, 
it is imperative to understand the evolution of small scales structure and all physical effects that contribute to this evolution. Here we discuss the 
possible implications and next steps in exploring the relative velocity effect.

Early galaxies may be observed in the next decade, either directly via the James Webb Space Telescope or indirectly through the near-infrared background
and its fluctuations.  Due to the relative velocity effect, the large-scale clustering of these galaxies
should show scale-dependent bias and (if sufficient statistics are available to split the galaxy population into multiple samples)
relative stochasticity between different samples of galaxies.  
For example, in our calculation at $z=15$, haloes of mass $10^8\,M_\odot$ show an increase of $\Delta b=0.73$ over the no-velocity result of $b_0=4.6$
at $k=0.02\,$Mpc$^{-1}$.
Whether this effect will be detectable depends on the as-yet-unknown luminosity function
of the highest-redshift galaxies, and whether the relative velocity effect can be separated from the scale-dependent bias produced by
reionization \cite{2006MNRAS.365.1012F, 2007MNRAS.377.1175D}.

Since reionization is believed to be driven by the formation of the first haloes massive enough to contain gas and produce stellar photoionizing 
radiation \cite{2001PhR...349..125B}, the relative velocity effect will delay reionization.  However, this effect is degenerate with the mapping from 
halo mass to the number of massive stars formed and given the modest (factor of $\sim 2$) effects investigated here we do not expect that the effects 
can be disentangled from the reionization history alone.  More interesting would be an investigation of the spatial structure of reionization and of 
the related high-redshift 21 cm signal, which has been investigated analytically and in simulations \cite{2003MNRAS.341...81I, 2004ApJ...608..622Z, 
2004ApJ...613....1F, 2006MNRAS.365..115F, 2006MNRAS.366..689C, 2007ApJ...654...12Z}.  The scale-dependent bias and stochasticity we find here for
early haloes may have a significant effect on the structure (power spectrum, topology, and correlation with the matter density field) of early
reionization bubbles.

If the {\em pre}-reionization 21 cm signal \cite{2004PhRvL..92u1301L} is ever observed, and cosmological information
extracted, the relative velocity effect will be very important: the smallest scale fluctuations in the baryons are modified at the tens
of percents level.  Indeed, since the 21 cm signal is nonlinear in the baryon density (in the limit where the hydrogen spin temperature is
closely coupled to the CMB temperature, the signal is proportional to $n^2$ times the temperature-dependent collision cross section \cite{2005ApJ...622.1356Z}), it is likely that even the
large-scale fluctuations would be affected because regions with increased small-scale baryon power spectra will show more absorption.
The locally anisotropic nature of the small-scale baryonic perturbations would also represent an issue for weak lensing of the 21 cm field \cite{2004NewA....9..417P, 2006ApJ...647..719M, 
2006ApJ...653..922Z, 2007MNRAS.381..447M} and/or 
non-Gaussianity searches 
\cite{2006PhRvL..97z1301C, 2008PhRvD..77j3506C}.
A full analysis of the effect on the 21 cm power spectrum and non-Gaussian statistics is deferred to future work.

Finally, the $\Lambda$CDM cosmology predicts that early dark matter haloes in the affected range of scales (mainly $\sim 10^4$--$10^8\,M_\odot$) are
assimilated into larger structures.  Some of these early haloes may still be present today as dark matter substructure, which has attracted a great
deal of interest since the subhalo mass function is in principle sensitive to the primordial small-scale CDM power spectrum and hence to possible
deviations from ``vanilla'' CDM behavior (e.g. warm dark matter, or particles that are kinetically coupled to baryons at high redshift).
Unfortunately, the overall power suppression effect we describe is probably {\em not} detectable via substructure since the transition from an initial 
CDM power spectrum through the formation and survival of substructure is still not quantitatively understood (e.g. \cite{2003ApJ...598...49Z, 
2007ApJ...667..859D, 2008MNRAS.391.1685S}).  However, the 
power suppression effect is modulated by 
the relative velocity field, which comes primarily from large-scale modes in the primordial density fluctuations ($k\sim 0.1\,$Mpc$^{-1}$) and hence can be reconstructed from large-scale structure surveys. Therefore it would be valuable for future work to investigate whether the $v_{\rm bc}$ effect can be detected by differential measurements that compare the substructure abundance in strong lens systems where the lens haloes have similar mass but different reconstructed ${\bf v}_{\rm bc}$. It is also important to mention that the suppression of the formation of the first halos, which seed present day galaxies, and the decrease in the high-k power spectrum might help alleviating the known problem of the over-abundance of substructure of dark matter halos in the $\Lambda$CDM model (the missing satellite problem). This effect might also be important for predictions of the annihilation signal from dark matter particles. We relegate detailed investigation of these questions to the future study.

In summary, we have shown that in the post-recombination Universe, there are bulk relative motions between the baryons and dark matter that are supersonic and are coherent over scales of several 
comoving megaparsecs.  The combined growth of small scale structure (between the baryon Jeans length $\lambda_{\rm J}$ and ${\cal M}\lambda_{\rm J}$, where ${\cal M}$ is the Mach number of relative 
motion) is suppressed due to the baryons advecting out of the potential wells created by the dark matter.  We find at lower redshifts (e.g. $z\sim 40$) a suppression of the power spectrum by $\sim 10$\% 
on scales of 50--500$\,$Mpc$^{-1}$ that is highly spatially variable.  The suppression results in some reduction in the abundance of early haloes, but more importantly changes their spatial structure, 
leading to scale-dependent bias and stochasticity of the first haloes.  These latter effects may be large for early low-mass, high-bias haloes, e.g. we find squared correlation coefficients
$\chi=r_{\rm hm}^2$ as small as $\sim 0.2$ at $z=40$.  Whether this unusual clustering pattern affects the spatial morphology of reionization depends on the importance of low-mass haloes and 
hence is unknown at this time, although we note that future 21 cm observations combined with simulations that distribute their sources of ionizing radiation in different ways may shed some light on this 
issue.  Farther in the future, the power suppression effect would certainly be significant for the interpretation of any pre-reionization 21 cm signal.  In any case, our analysis highlights the 
importance of reconsidering ``standard'' notions of structure formation (e.g. linear bias of haloes on large scales) as we enter new physical regimes at high redshift.

\section*{Acknowledgments}

We thank Neal Dalal, Eiichiro Komatsu, Kiyoshi Masui, Leonidas Moustakas, Michael Kuhlen and Ue-Li Pen for helpful conversations.

D.T. and C.H. are supported by the U.S. Department of Energy (DE-FG03-92-ER40701) and the National Science Foundation (AST-0807337).
C.H. is supported by the Alfred P. Sloan Foundation.


\bibliography{VbdPaper}

\end{document}